\documentclass{article}
\pdfoutput=1
    \PassOptionsToPackage{numbers, compress}{natbib}
\usepackage[preprint]{neurips_2023}

\usepackage[utf8]{inputenc} % allow utf-8 input
\usepackage[T1]{fontenc}    % use 8-bit T1 fonts
\usepackage{hyperref}       % hyperlinks
\usepackage{url}            % simple URL typesetting
\usepackage{booktabs}       % professional-quality tables
\usepackage{amsfonts}       % blackboard math symbols
\usepackage{nicefrac}       % compact symbols for 1/2, etc.
\usepackage{microtype}      % microtypography
\usepackage{xcolor}         % colors
\usepackage{graphicx}
\usepackage{amsmath}
\usepackage{wrapfig}
\usepackage{caption}
\usepackage{subcaption}
\usepackage{listings}
\usepackage{listings,lstautogobble}
\DeclareMathOperator*{\argmax}{\arg\max}

\title{Symbolic Learning for Material Discovery}

\author{Dan Cunnington \\
        IBM Research Europe \\
        \texttt{dancunnington@uk.ibm.com} 
        \And
        Flaviu Cipcigan \\
        IBM Research Europe \\
        \texttt{flaviu.cipcigan@ibm.com} 
        \And
        Rodrigo Neumann Barros Ferreira \\
        IBM Research Brazil\\
        \texttt{rneumann@br.ibm.com} \\
        \And
        Jonathan Booth \\
        Science and Technology Facilities Council \\
        \texttt{jonathan.booth@stfc.ac.uk}
        }

\begin{document}

\maketitle

\begin{abstract}
  % It is of paramount importance to accelerate the discovery of new materials in order to meet climate change targets, as evaluating a candidate material for a given task is often computationally expensive. Recent methods aim to discover a near-optimal material in a given database without a brute-force evaluation. However, these methods lack the interpretability required in order for domain experts to inspect learned models and verify physical and chemical soundness. In this paper, we introduce SyMDis, a sample efficient optimisation method that learns interpretable rules to inform material selection. We demonstrate SyMDis is able to discover near optimal materials in large databases with few samples, performing comparably to a state-of-the-art optimiser, whilst learning interpretable rules out-of-the-box. Furthermore, the rules learned by SyMDis can generalise to unseen datasets and return high performing candidates in a zero-shot evaluation, which would be difficult to achieve with other approaches.

 Discovering new materials is essential to solve challenges in climate change, sustainability and healthcare. A typical task in materials discovery is to search for a material in a database which maximises the value of a function. That function is often expensive to evaluate, and can rely upon a simulation or an experiment. Here, we introduce SyMDis, a sample efficient optimisation method based on symbolic learning, that discovers near-optimal materials in a large database. SyMDis performs comparably to a state-of-the-art optimiser, whilst learning interpretable rules to aid physical and chemical verification. Furthermore, the rules learned by SyMDis generalise to unseen datasets and return high performing candidates in a zero-shot evaluation, which is difficult to achieve with other approaches.
\end{abstract}

\section{Introduction}
Accelerating the discovery of new materials is one of the fundamental challenges facing the scientific community. Many important applications such as carbon capture and battery technology are reliant on new materials in order to realise their potential value and limit global warming in accordance with the 2015 Paris agreement \cite{paris}. However, evaluating the large design landscape is a computationally infeasible task, and there is simply not enough time to proceed using traditional methods. Therefore, there is increasing attention from the AI community to accelerate material discovery \cite{LI2020393, LIU2017159}.

Many approaches aim to discover an optimal material within a given database by running in-silico simulations to estimate a desired metric. Whilst the naive brute-force approach is obviously inefficient, optimisation techniques such as Bayesian Optimisation and active learning select high performing candidates whilst minimising the number of evaluations of an expensive in-silico simulation \cite{D1ME00154J,lookman2019active}. This is often achieved by balancing an exploration vs. exploitation trade-off of the material search space. However, the underlying machine learning models used are often difficult to interpret, and also can not be easily transferred to new datasets. Therefore, it is unclear whether materials are being selected based on criteria that are physically and chemically sound.

To tackle this problem, we introduce \textbf{SyMDis}: Symbolic learning for Material Discovery, which exploits symbolic AI techniques to learn naturally interpretable rules that map material descriptors to the desired performance metric. SyMDis is inspired by active learning, and on each iteration, a small number of samples for in-silico computation are selected from a database based on the learned rules. Then, new rules are learned, and a new batch of materials are selected for evaluation. The goal is to discover a high performing material within a large database, whilst minimising the number of calls to the (expensive) in-silico computation. We evaluate SyMDis for the task of identifying suitable Metal Organic Frameworks (MOFs) to maximise $\mathrm{CO_2}$ uptake for carbon capture. Our experiments show that SyMDis performs comparably to a state-of-the-art Bayesian Optimisation method, returning a MOF with a $\mathrm{CO_2}$ uptake within 92.5\% of the maximum in a database of 19.4K MOFs, after only 100 calls to the objective function. This increases to a MOF with 97.5\% of the maximum with 250 calls. Crucially, SyMDis learns interpretable rules that express why certain MOFs were chosen in terms of various MOF descriptors. This enables human domain experts to verify the learned rules are physically and chemically sound. Furthermore, our evaluation demonstrates the rules learned by SyMDis can generalise to new unseen datasets in a zero-shot evaluation, obtaining high performing MOFs without requiring \textit{any} calls to the objective function on the new dataset. This would be difficult to achieve with other approaches. %The remainder of the paper is organised as follows. We introduce SyMDis in Section \ref{sec:method}, present our experimental results on the MOF domain in Section \ref{sec:results}, and discuss related work in Section \ref{sec:related_work}, as well as explaining how SyMDis could be used as part of a workflow with generative models. Finally, we provide concluding remarks in Section \ref{sec:conclusion}. 

% Maybe drop the link to generative models if low on space. Or move to appendix.

\section{Method}\label{sec:method}
\textbf{Problem statement}. Let us assume a material is represented by a set of descriptor value pairs $\boldsymbol{x} = \{\langle d,v\rangle,\ldots \}$, where $d\in\cal{D}$ is a descriptor and $v\in\cal{V}$ is an associated value. In this paper, we assume each descriptor has a defined value, i.e., $\forall d\in \cal{D}$, there exists a $v^{\prime}\in\cal{V}$ s.t. $\langle d, v^{\prime}\rangle\in\boldsymbol{x}$. An objective function $f: (\cal{D}\times\cal{V})^{\vert\cal{D}\vert} \rightarrow \mathbb{R}$ maps a set of descriptor value pairs into a target metric $y \in \mathbb{R}$. Let us also assume a database of \textit{unlabelled} material samples $B = \{ \boldsymbol{x},\ldots\}$. The goal is to find the material in $B$ that maximises $y$, given $f$, i.e., $\boldsymbol{x}^{*} = \argmax_{\boldsymbol{x}\in B} f(\boldsymbol{x})$, whilst minimising the number of evaluations of $f$, as $f$ could be computationally expensive (e.g., an in-silico simulation). To tackle this problem, SyMDis uses an iterative approach, inspired by active learning.

\begin{wrapfigure}{r}{0.5\textwidth}
    \centering
    \includegraphics[width=0.5\textwidth]{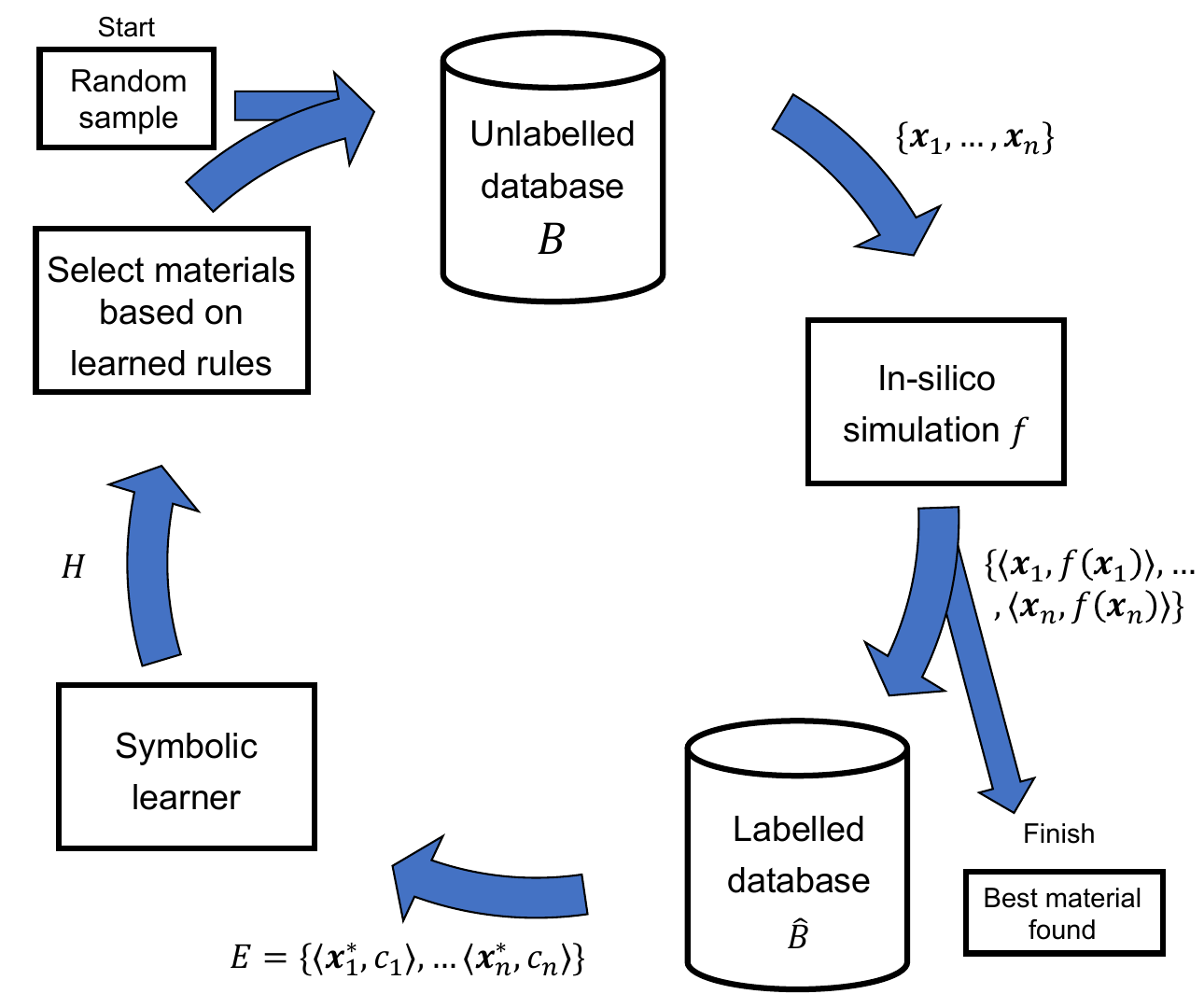}
    \caption{SyMDis architecture}
    \label{fig:arch}
\end{wrapfigure}

\textbf{SyMDis}. In addition to the database of unlabelled materials $B$, let us also define a database of (initially empty) \textit{labelled} materials, $\hat{B}=\emptyset$. SyMDis begins with a random sample of $n$ materials $\{\boldsymbol{x}_{1},...,\boldsymbol{x}_{n} \}$ chosen from $B$, which are evaluated using $f$, to obtain a set of targets $\{f(\boldsymbol{x}_{1}),...,f(\boldsymbol{x}_{n}) \}$. These materials are then removed from $B$, and alongside their targets, are stored in $\hat{B}$, i.e., $\hat{B} = \hat{B}\cup \{ \langle \boldsymbol{x}_1,f(\boldsymbol{x}_1)\rangle,..,\langle\boldsymbol{x}_n,f(\boldsymbol{x}_n)\rangle \}$. SyMDis then constructs a set of training examples for a symbolic learner, which learns a set of logical rules that selects materials from $B$ for the next iteration. SyMDis terminates after a given number of iterations, or when $B$ is empty. The architecture is presented in Figure \ref{fig:arch}. Let us now describe how the labelled materials in $\hat{B}$ are converted into training examples for the symbolic learner.

\textbf{Generating symbolic training examples}. On each iteration, a set $\{\langle\boldsymbol{x}^{*}_{1},y^{*}_{1}\rangle,...,\langle\boldsymbol{x}^{*}_{n},y^{*}_{n}\rangle \}$ of $n$ samples with the largest target metrics $y^{*}$ are selected from $\hat{B}$. Note on the first iteration this set of samples is equal to the initial random samples. SyMDis then generates a classification for each sample, by binning the $y^{*}$ values into a class label $c \in \{\text{\textit{excellent}}, \text{\textit{good}}, \text{\textit{moderate}}, \text{\textit{poor}} \}$. The samples with a $y^{*}$ value in the largest 10\% receive the label $c=\text{\textit{excellent}}$, the next 20\% receive the label $c=\text{\textit{good}}$, the next 30\% receive the label $c=\text{\textit{moderate}}$, and the remaining 40\% receive the label $c=\text{\textit{poor}}$. Generating class labels in this manner helps to prevent over-fitting to a particular sample, as provided $n$ is large enough, multiple samples receive the same class label. This particular percentage split was chosen to enable the unlabelled database to be filtered based on rules learned from high performing candidates, i.e., candidates within the top 10\%, as priority is given to \textit{excellent} candidates during filtering for the next iteration, see \textit{Selecting materials} below. The chosen split performs well in our experiments, and further tuning and investigation is left as future work. The set of examples for the symbolic learner can then be generated as $E=\{\langle\boldsymbol{x}^{*}_{1},c_{1}\rangle,...,\langle\boldsymbol{x}^{*}_{n},c_{n}\rangle \}$.

\textbf{Symbolic learner}. A symbolic learner can then learn a logic program called a hypothesis $H\in\cal{H}$ that explains the training examples. In this paper, we assume $H$ is a propositional logic program, although since the symbolic learning component in SyMDis is modular, this program could be of any logical expressivity or in any logical format. Given this modularity, we now define a general notion of a symbolic learner. Firstly, the score of a hypothesis w.r.t. a training example $\langle\boldsymbol{x}^{*},c\rangle$ is defined as:

\begin{equation}
    SCORE(H,\langle\boldsymbol{x}^{*},c\rangle) = \begin{cases} 1 &\text{if}\  H(\boldsymbol{x}^{*}) \models c\\
    0 &\text{otherwise}
    \end{cases}
\label{eqn:score}
\end{equation}
where $\models$ denotes logical entailment. The goal of the symbolic learner is to learn a hypothesis $H$ s.t.:
\begin{equation}
    H^{*} = \argmax_{H\in\cal{H}} \sum_{\langle\boldsymbol{x}^{*},c\rangle \in E} SCORE(H, \langle\boldsymbol{x}^{*},c\rangle)
\end{equation}

which intuitively means to learn a $H$ that maximises the number of training examples where the correct class label is output, given the input descriptors.

\textbf{Selecting materials} Finally, SyMDis selects a new batch of $n$ samples from $B$ using the learned $H$. This is achieved by evaluating all of the materials in $B$ using $H$ to determine a predicted class label $\hat{c}\in \{\text{\textit{excellent}}, \text{\textit{good}}, \text{\textit{moderate}}, \text{\textit{poor}} \}$. Relative to $f$, this is a cheap computation. SyMDis progressively selects samples in the order of the quality of the predicted label (\textit{excellent} $\rightarrow$ \textit{poor}), until a total of $n$ is reached. A random sample is performed if selecting all samples within a class would exceed the total of $n$. This forms the new batch of samples
$\{\boldsymbol{x}_1,\ldots,\boldsymbol{x}_{n}\}$ for the next iteration. %Selecting materials in this manner enables SyMDis to converge to more optimal materials, as the class labels in the learned $H$ are defined in terms of better target values on subsequent iterations.
% todo: algorithm
% 
\section{Experiments}\label{sec:results}
To evaluate SyMDis, we use three MOF datasets, where the task is to discover the best material for performing carbon capture, whilst minimising the number of in-silico adsorption simulations required. The target metric is to maximise \textit{working capacity}, which is defined as the amount of $\mathrm{CO_2}$ adsorbed during high pressure minus the remaining $\mathrm{CO_2}$ left on the material during desorption (low pressure). Our evaluation aims to address the following questions; (\textbf{Q1}) How does SyMDis compare to a state-of-the-art Bayesian Optimisation approach, known for its sample efficiency? (\textbf{Q2}) Can SyMDIS take advantage of a logic-based symbolic learner to improve performance and/or interpretability? (\textbf{Q3}) Can SyMDis discover physically and chemically sound rules? (\textbf{Q4}) Can the learned rules generalise to unseen data, to discover a high performing MOF without requiring \textit{any} calls to the objective function on the new dataset (zero-shot)?

% TODO DAN: Explain why working capacity requires high and low pressure - an adsorption swing?? Ask Flaviu/Rodrigo to contribute. Need to explain in simple terms. Maybe I can google.. 

\textbf{Setup}. We utilise three MOF datasets from \cite{moosavi2020understanding}; \textit{ARABG}, \textit{CoRE2019}, and \textit{BW20K} which contain 387, 9525, and 19,379 MOFs respectively. These datasets are annotated with $\mathrm{CO_2}$ adsorption at pressures 0.15bar and 16bar, with a fixed temperature of 298K. Note that 0.15bar used by \cite{moosavi2020understanding} is an allusion to the partial pressure of $\mathrm{CO_2}$ in coal-fired flue gas (15\% $\mathrm{CO_2}$ / 85\% $\mathrm{N_2}$) at ambient pressure. These annotations enable us to perform a comprehensive evaluation without running the simulations, i.e., our objective function is simply a look-up in the database for these features, which can be used to calculate the working capacity. No modifications to the datasets are performed, other than selecting 15 descriptors based on a regression analysis to determine feature importance (see Table \ref{tab:desc} in Appendix \ref{sec:appendix:learned_rules}). We generate 20 random seeds and select 20 sets of 50 materials at random from each dataset. We plot the best working capacity achieved in terms of a percentage of the maximum, w.r.t. the number of evaluations of our objective function, averaged over the 20 repeats. 

For the symbolic learning component, we use a Decision Tree from scikit-learn (denoted \textbf{SyMDis DT}), and to address evaluation Q2, FastLAS \cite{law2020fastlas} (denoted \textbf{SyMDis FastLAS}), a recent logic-based machine learning approach known for it's ability to generalise and learn expressive logic programs in the form of Answer Set Programming \cite{gelfond2014knowledge}. FastLAS can also learn from noisy data, where each example is given a weight, and the symbolic learner is encouraged to cover examples with higher weight. In our experiments, we assign higher weights to better performing materials to encourage the symbolic learner to learn rules that cover the high performing candidates. We use weights 10, 5, 2, and 1 for the classes excellent, good, mediocre and poor respectively. Future work could investigate additional methods for selecting the weights, and use them to express uncertainty on the descriptor values. We evaluate each variant with varying batch sizes $n \in \{ 10, 20,..., 60\}$ and present the best performing result. We also note that FastLAS timed-out after 1 minute with a batch size $n > 30$, so the SyMDis FastLAS variant uses a batch size of $n=30$ for all experiments.

\textbf{Baselines}. To investigate evaluation Q1, we compare to IBM's state-of-the-art Bayesian Optimisation library (\textbf{BOA}) \cite{jasrasaria2019dynamic}. Following consultation with the authors, and our own experiments, we use Expected Improvement with a Basic Sampler, the Adaptive Epsilon setting, enable $x$ and $y$ normalisation, and set the batch size to 1, as these configurations resulted in the best performance. We also evaluate a naive random baseline (\textbf{Random}), that randomly selects a material from $B$ at each iteration. For both baselines, we initialise using the same 20 sets of random materials as SyMDis.

\begin{figure}[t]
     \centering
     \begin{subfigure}[b]{0.32\textwidth}
         \centering
         \includegraphics[width=\textwidth]{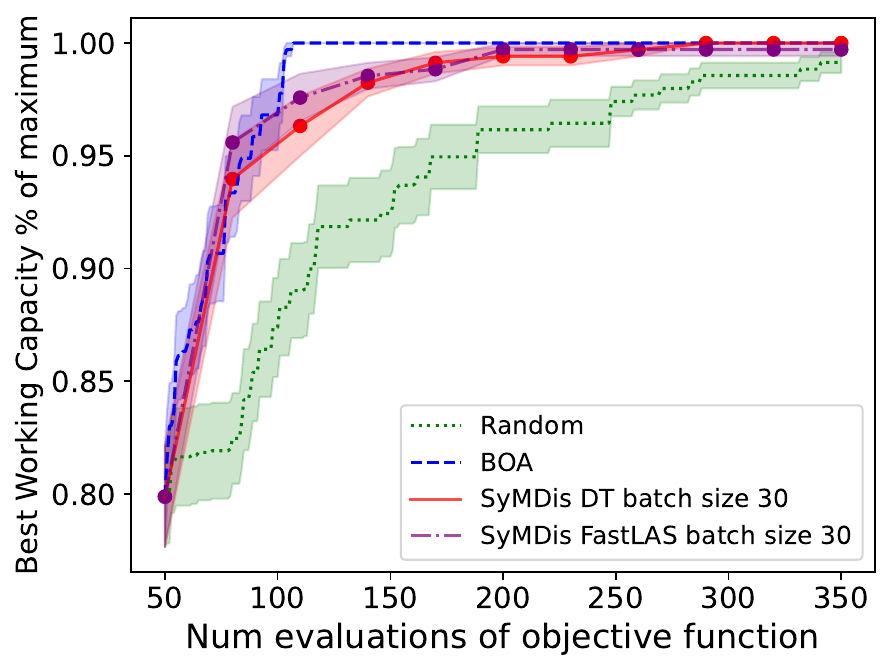}
        \caption{ARABG}
         \label{fig:arabg}
     \end{subfigure}
     \hfill
     \begin{subfigure}[b]{0.32\textwidth}
         \centering
         \includegraphics[width=\textwidth]{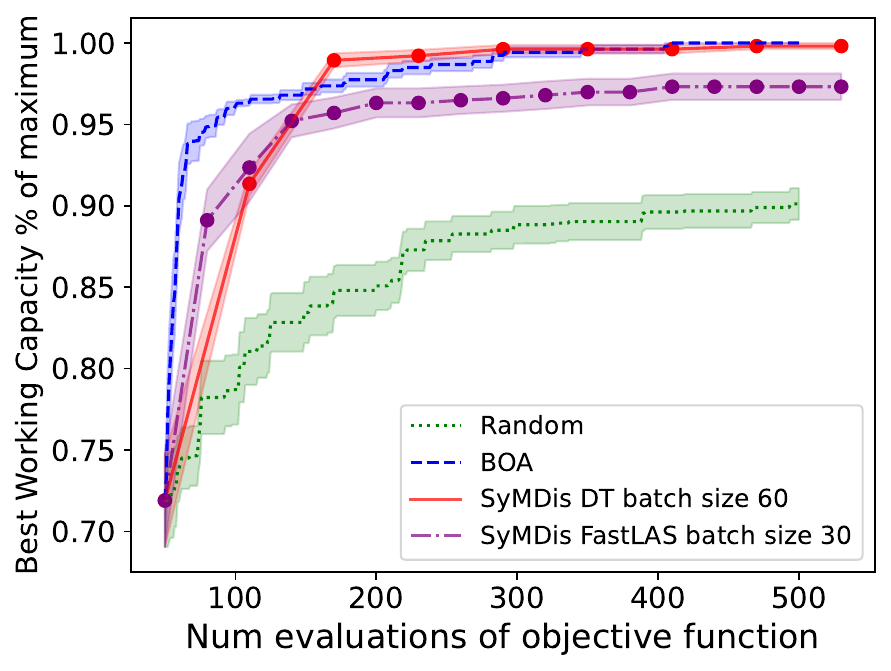}
         \caption{CoRE2019}
         \label{fig:core2019}
     \end{subfigure}
     \hfill
     \begin{subfigure}[b]{0.32\textwidth}
         \centering
         \includegraphics[width=\textwidth]{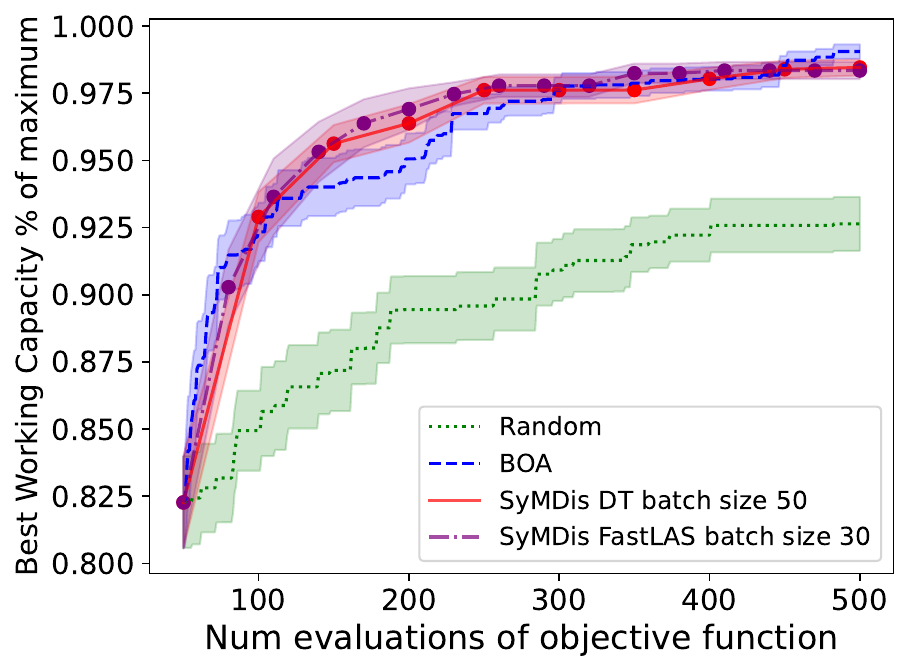}
         \caption{BW20K}
         \label{fig:bw20k}
     \end{subfigure}
        \caption{MOF results. Shaded regions indicate standard error.}
        \label{fig:results}
\end{figure}

\begin{figure}[t]
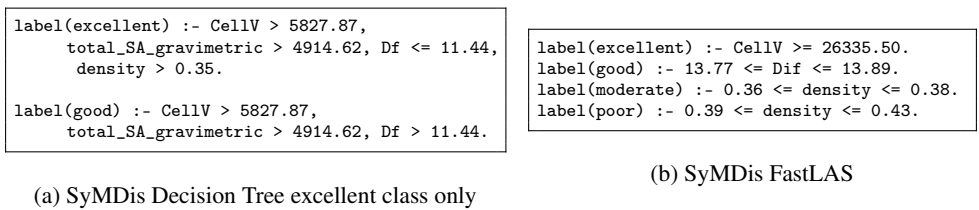

    \centering
    \begin{subfigure}[m]{0.46\textwidth}
        \begin{lstlisting}
label(excellent) :- CellV > 5827.87, total_SA_gravimetric > 4914.62, Df <= 11.44, density > 0.35.

label(good) :- CellV > 5827.87, total_SA_gravimetric > 4914.62, Df > 11.44.
\end{lstlisting}
    \caption{SyMDis Decision Tree excellent class only}
    \label{fig:dt_rules}
    \end{subfigure}
    \hspace{10pt}
    \begin{subfigure}[m]{0.41\textwidth}
        \begin{lstlisting}
label(excellent) :- CellV >= 26335.50.
label(good) :- 13.77 <= Dif <= 13.89.
label(moderate) :- 0.36 <= density <= 0.38.
label(poor) :- 0.39 <= density <= 0.43.
    \end{lstlisting}    
        \caption{SyMDis FastLAS}
        \label{fig:fastlas_rules}
    \end{subfigure}
    \caption{Example rules learned on BW20K}
    \label{fig:learned_rules}
\end{figure}

\begin{figure}[t]
    \centering
    \begin{minipage}{.35\textwidth}
      \centering
      \includegraphics[width=0.99\textwidth]{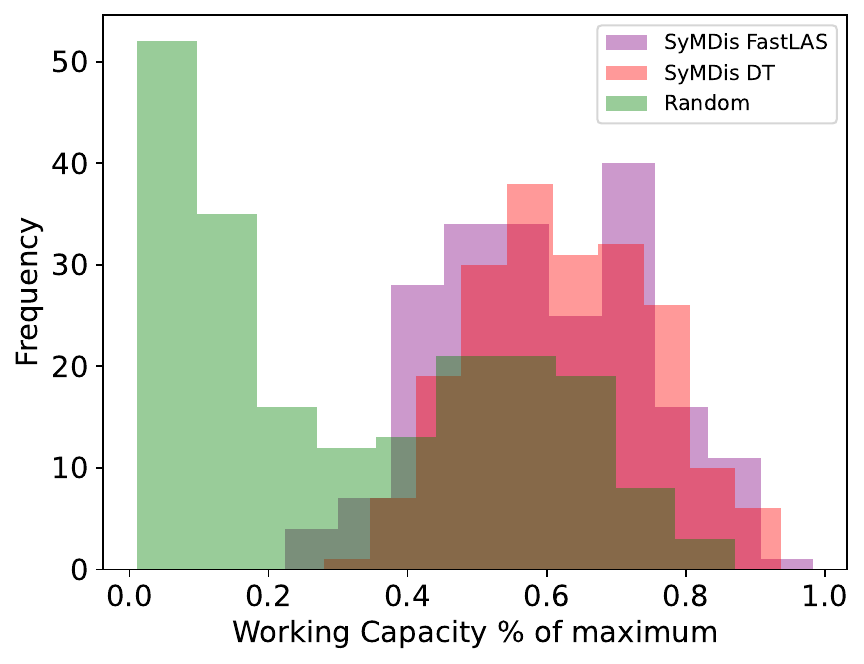}
      \captionof{figure}{Distribution of MOFs sampled from BW20K using rules learned on ARABG}
      \label{fig:gen_ex}
    \end{minipage}%
    \hspace{2em}
    \begin{minipage}{.54\textwidth}
    \begin{subfigure}[t]{0.49\textwidth}
        \centering
        \includegraphics[width=\textwidth]{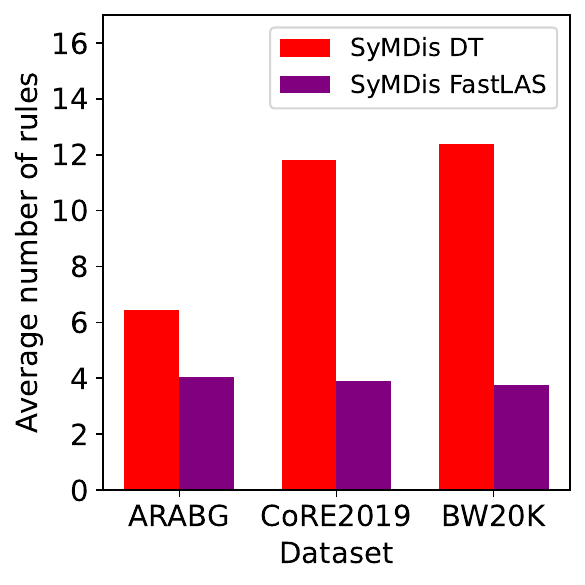}
        \caption{Avg. number of rules}
        \label{fig:num_rules}
    \end{subfigure}
    \begin{subfigure}[t]{0.49\textwidth}
        \centering
        \includegraphics[width=\textwidth]{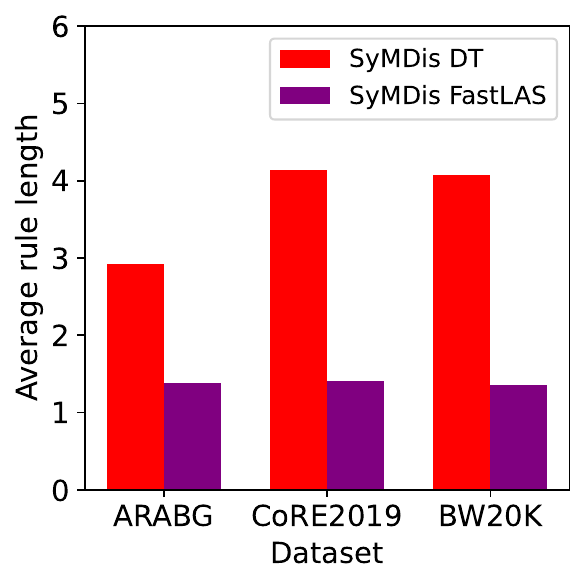}
        \caption{Avg. rule length}
        \label{fig:avg_len}
    \end{subfigure}
    \captionof{figure}{Interpretability of the Decision Tree and FastLAS learned rules.}
    \label{fig:rule_len_analysis}
    \end{minipage}
\end{figure}

\textbf{Results}. The results are presented in Figure \ref{fig:results}, and example rules are shown in Figure \ref{fig:learned_rules}, for one repeat on the BW20K dataset. SyMDis discovers a near optimal MOF in all cases, significantly outperforming random selection. On BW20K, the most challenging dataset, SyMDis discovers a MOF with 92.5\% of the maximal working capacity whilst only evaluating the objective function for 100 MOFs out of a total of $\sim$20K samples. This increases to 97.5\% with 250 evaluations. To answer evaluation Q1, SyMDis performs comparably to BOA on all datasets, whilst learning naturally interpretable rules (i.e., no post-hoc interpretability is required). For evaluation Q2, FastLAS does not appear to offer an advantage when compared to a Decision Tree in terms of performance, although learns a significantly shorter set of rules, which are easier to interpret (see Figure \ref{fig:rule_len_analysis} and Appendix \ref{sec:appendix:learned_rules} for analysis). The rules learned by SyMDis enable manual inspection, and indeed correspond to a reasonable chemical explanation. For example, the rules learned on the BW20K dataset for both SyMDis variants show that the unit cell volume descriptor is important. The unit cell within the MOF is the structure that is repeated to make the crystal. It’s possible that a larger unit cell is likely to have more space, which could influence adsorption \cite{guo2015zeolite}. Also, \cite{fernandez} reports that the gravimetric surface area, pore size, and void fraction are important geometric descriptors for predicting $\mathrm{CO_2}$ uptake. As you can see in Figure \ref{fig:learned_rules}, the SyMDis Decision Tree variant also uses the gravimetric surface area descriptor. This answers evaluation Q3.

% \begin{wrapfigure}{r}{0.3\textwidth}
% \vspace{-1em}
%     \centering
%     \includegraphics[width=0.3\textwidth]{figures/gen_ex.pdf}
%     \caption{Distribution of MOFs sampled from BW20K using rules learned on ARABG.}
%     \label{fig:gen_ex}
% \end{wrapfigure}

To address Q4, we perform a zero-shot transfer, and apply the rules learned on a source dataset to a target dataset which is completely unseen during the optimisation. We use the learned rules to obtain a predicted class $\hat{c}$ for every material in the target dataset, and take a random sample of 200 materials predicted as \textit{excellent}. We then plot the distribution of working capacities from the sampled MOFs, and compare to 200 random samples. We evaluate all combinations of source/target datasets except where the target is ARABG, as this dataset only contains 387 MOFs. The full results are shown in Appendix \ref{sec:app:generalisation}, with an example shown in Figure \ref{fig:gen_ex}. In most cases, the distribution of working capacity values calculated from the sampled materials is larger than those calculated from a random selection. Surprisingly, even the rules learned on the ARABG dataset can generalise to CoRE2019 and BW20K which have a significantly larger number of MOFs.

Finally, in terms of run-time, SyMDis is very efficient. The average wall-clock time for the Decision Tree variant is 0.11, 2.43, 4.96 seconds to complete the optimisation in full for the ARABG, CoRE2019, and BW20K datasets respectively, and 10.58, 25.25, and 52.57 seconds for the FastLAS variant. Note these run-times include the database lookup as our objective function. In reality, SyMDis adds minimal overhead compared to an expensive objective function.

% TODO: Increase font size in x/y axis labels / ticks.

\section{Related work}\label{sec:related_work}
SyMDis is inspired by active learning which interactively queries a third-party to label data \cite{settles2009active}. Bayesian Optimisation (BO) \cite{D1ME00154J,snoek2012practical} is a popular method of active learning, where a surrogate model of an objective function is iteratively constructed given data observations. Whilst BO has achieved state-of-the-art performance in many applications \cite{D1ME00093D,shahriari2015taking}, the surrogate model is often a black-box which requires additional components to explain what's been learned. We have demonstrated SyMDis performs comparably to BO on the MOF datasets, whilst providing naturally interpretable rules. A recent approach also learns surrogate MOF adsorption models \cite{mukherjee2023active}, but relies on a black-box Gaussian Process Regressor. Many approaches have been developed to aid with interpretability, particularly in material discovery \cite{imlm,fernandez,mikulskis2019toward,muckley2023interpretable}. However, these approaches rely on statistical machine learning techniques such as Decision Trees, Random Forests, and Linear Models, which as shown in Figure \ref{fig:rule_len_analysis} and Appendix \ref{sec:appendix:learned_rules}, can become difficult to interpret when applied to large datasets. In contrast, SyMDis can take advantage of logic-based techniques which offer improved interpretability. As discussed in Section \ref{sec:results}, \cite{fernandez} uses a Decision Tree to learn rules for $\mathrm{CO_2}$ uptake, using MOFs from the Northwestern University database \cite{wilmer2012large}. SyMDis learns rules using similar descriptors, and extends this work by integrating the symbolic learner into an active learning loop.

In the context of scientific discovery, symbolic techniques have been applied widely for the purposes of symbolic regression \cite{burlacu2023symbolic,fitzsimmons2018symbolic,wang_wagner_rondinelli_2019,weng2020simple}, where the goal is to generate a fully interpretable mathematical expression that fits a set of training data points. However, learning a general expression is challenging, and these methods are often sensitive to noise \cite{udrescu2020ai}. In contrast, instead of trying to learn an exact mathematical expression, SyMDis learns logical rules that can take advantage of descriptors with different data types (i.e., non-numerical categorical descriptors), and generalise more easily over a set of noisy examples. Furthermore, using a logic-based symbolic learner, SyMDis can easily include existing background knowledge, ensure constraints are satisfied, and learn various logical relations and comparison operators. %Finally, please refer to Appendix \ref{sec:app:generative_models} for a forward-thinking discussion on how SyMDis relates to recent efforts around generative models.

% MOF specific

% Refer to appendix C for generative models
\section{Conclusion}\label{sec:conclusion}
We have introduced \textbf{SyMDis}, a sample-efficient optimisation method that discovers near-optimal materials in a given database, whilst minimising the number of calls to an objective function. We have applied SyMDis to a MOF use-case, and demonstrated SyMDis is able to discover a MOF for $\mathrm{CO_2}$ uptake with 97.5\% of the maximum in a database with 19.4K MOFs, requiring only 250 calls to the objective function. Furthermore, SyMDis learns interpretable rules to aid chemical and physical verification, and can generalise via a zero-shot transfer to unseen target datasets. 

\textbf{Future Work}. One could extend SyMDis to multi-objective optimisation problems by designing a target objective that incorporates multiple metrics (e.g., \textit{target objective} = \textit{working capacity} + \textit{selectivity} - \textit{cost}. One could also apply SyMDis in other domains, such as solvents or additives. Finally, one could integrate SyMDis with generative models such as GFlowNets \cite{bengio2021flow}, by using SyMDis to discover the most optimal candidate output from a generative model. This would combine the benefits of both generation and optimisation approaches to material discovery, as SyMDis could apply more thorough evaluations of novel candidates, before handing off to a lab for further analysis.

\section*{Acknowledgments}
The authors wish to acknowledge Edward Pyzer-Knapp and Clyde Fare from IBM Research for helping to identify the appropriate BOA configuration for us to use in our experiments. This work was funded through the UKRI Hartree National Centre for Digital Innovation.

\bibliographystyle{plain}
\bibliography{ref}

\begin{thebibliography}{10}

\bibitem{imlm}
Alice E.~A. Allen and Alexandre Tkatchenko.
\newblock Machine learning of material properties: Predictive and interpretable multilinear models.
\newblock {\em Science Advances}, 8(18):eabm7185, 2022.

\bibitem{bengio2021flow}
Emmanuel Bengio, Moksh Jain, Maksym Korablyov, Doina Precup, and Yoshua Bengio.
\newblock Flow network based generative models for non-iterative diverse candidate generation.
\newblock {\em Advances in Neural Information Processing Systems}, 34:27381--27394, 2021.

\bibitem{burlacu2023symbolic}
Bogdan Burlacu, Michael Kommenda, Gabriel Kronberger, Stephan~M Winkler, and Michael Affenzeller.
\newblock Symbolic regression in materials science: Discovering interatomic potentials from data.
\newblock In {\em Genetic Programming Theory and Practice XIX}, pages 1--30. Springer, 2023.

\bibitem{D1ME00093D}
Aryan Deshwal, Cory~M. Simon, and Janardhan~Rao Doppa.
\newblock Bayesian optimization of nanoporous materials.
\newblock {\em Mol. Syst. Des. Eng.}, 6:1066--1086, 2021.

\bibitem{D1ME00154J}
Sanket Diwale, Maximilian~K. Eisner, Corinne Carpenter, Weike Sun, Gregory~C. Rutledge, and Richard~D. Braatz.
\newblock Bayesian optimization for material discovery processes with noise.
\newblock {\em Mol. Syst. Des. Eng.}, 7:622--636, 2022.

\bibitem{fernandez}
Michael Fernandez and Amanda~S. Barnard.
\newblock Geometrical properties can predict co2 and n2 adsorption performance of metal–organic frameworks (mofs) at low pressure.
\newblock {\em ACS Combinatorial Science}, 18(5):243--252, 2016.
\newblock PMID: 27022760.

\bibitem{fitzsimmons2018symbolic}
Jake Fitzsimmons and Pablo Moscato.
\newblock Symbolic regression modeling of drug responses.
\newblock In {\em 2018 First International Conference on Artificial Intelligence for Industries (AI4I)}, pages 52--59. IEEE, 2018.

\bibitem{gelfond2014knowledge}
Michael Gelfond and Yulia Kahl.
\newblock {\em Knowledge representation, reasoning, and the design of intelligent agents: The answer-set programming approach}.
\newblock Cambridge University Press, 2014.

\bibitem{guo2015zeolite}
Peng Guo, Jiho Shin, Alex~G Greenaway, Jung~Gi Min, Jie Su, Hyun~June Choi, Leifeng Liu, Paul~A Cox, Suk~Bong Hong, Paul~A Wright, et~al.
\newblock A zeolite family with expanding structural complexity and embedded isoreticular structures.
\newblock {\em Nature}, 524(7563):74--78, 2015.

\bibitem{jasrasaria2019dynamic}
Dipti Jasrasaria and Edward~O Pyzer-Knapp.
\newblock Dynamic control of explore/exploit trade-off in bayesian optimization.
\newblock In {\em Intelligent Computing: Proceedings of the 2018 Computing Conference, Volume 1}, pages 1--15. Springer, 2019.

\bibitem{law2020fastlas}
Mark Law, Alessandra Russo, Elisa Bertino, Krysia Broda, and Jorge Lobo.
\newblock Fastlas: Scalable inductive logic programming incorporating domain-specific optimisation criteria.
\newblock In {\em Proceedings of the AAAI conference on artificial intelligence}, volume~34, pages 2877--2885, 2020.

\bibitem{LI2020393}
Jiali Li, Kaizhuo Lim, Haitao Yang, Zekun Ren, Shreyaa Raghavan, Po-Yen Chen, Tonio Buonassisi, and Xiaonan Wang.
\newblock Ai applications through the whole life cycle of material discovery.
\newblock {\em Matter}, 3(2):393--432, 2020.

\bibitem{LIU2017159}
Yue Liu, Tianlu Zhao, Wangwei Ju, and Siqi Shi.
\newblock Materials discovery and design using machine learning.
\newblock {\em Journal of Materiomics}, 3(3):159--177, 2017.
\newblock High-throughput Experimental and Modeling Research toward Advanced Batteries.

\bibitem{lookman2019active}
Turab Lookman, Prasanna~V Balachandran, Dezhen Xue, and Ruihao Yuan.
\newblock Active learning in materials science with emphasis on adaptive sampling using uncertainties for targeted design.
\newblock {\em npj Computational Materials}, 5(1):21, 2019.

\bibitem{mikulskis2019toward}
Paulius Mikulskis, Morgan~R Alexander, and David~Alan Winkler.
\newblock Toward interpretable machine learning models for materials discovery.
\newblock {\em Advanced Intelligent Systems}, 1(8):1900045, 2019.

\bibitem{moosavi2020understanding}
Seyed~Mohamad Moosavi, Aditya Nandy, Kevin~Maik Jablonka, Daniele Ongari, Jon~Paul Janet, Peter~G Boyd, Yongjin Lee, Berend Smit, and Heather~J Kulik.
\newblock Understanding the diversity of the metal-organic framework ecosystem.
\newblock {\em Nature communications}, 11(1):1--10, 2020.

\bibitem{muckley2023interpretable}
Eric~S Muckley, James~E Saal, Bryce Meredig, Christopher~S Roper, and John~H Martin.
\newblock Interpretable models for extrapolation in scientific machine learning.
\newblock {\em Digital Discovery}, 2023.

\bibitem{mukherjee2023active}
Krishnendu Mukherjee, Etinosa Osaro, and Yamil~J Colon.
\newblock Active learning for efficient navigation of multi-component gas adsorption landscapes in a mof.
\newblock {\em Digital Discovery}, 2023.

\bibitem{paris}
United Nations.
\newblock Paris agreement 2015.
\newblock \url{https://treaties.un.org/pages/ViewDetails.aspx?src=TREATY&mtdsg_no=XXVII-7-d&chapter=27&clang=_en}.
\newblock Accessed: 30-08-2023.

\bibitem{settles2009active}
Burr Settles.
\newblock Active learning literature survey.
\newblock {\em University of Wisconsin-Madison Department of Computer Sciences}, 2009.

\bibitem{shahriari2015taking}
Bobak Shahriari, Kevin Swersky, Ziyu Wang, Ryan~P Adams, and Nando De~Freitas.
\newblock Taking the human out of the loop: A review of bayesian optimization.
\newblock {\em Proceedings of the IEEE}, 104(1):148--175, 2015.

\bibitem{snoek2012practical}
Jasper Snoek, Hugo Larochelle, and Ryan~P Adams.
\newblock Practical bayesian optimization of machine learning algorithms.
\newblock {\em Advances in neural information processing systems}, 25, 2012.

\bibitem{udrescu2020ai}
Silviu-Marian Udrescu and Max Tegmark.
\newblock Ai feynman: A physics-inspired method for symbolic regression.
\newblock {\em Science Advances}, 6(16):eaay2631, 2020.

\bibitem{wang_wagner_rondinelli_2019}
Yiqun Wang, Nicholas Wagner, and James~M. Rondinelli.
\newblock Symbolic regression in materials science.
\newblock {\em MRS Communications}, 9(3):793–805, 2019.

\bibitem{weng2020simple}
Baicheng Weng, Zhilong Song, Rilong Zhu, Qingyu Yan, Qingde Sun, Corey~G Grice, Yanfa Yan, and Wan-Jian Yin.
\newblock Simple descriptor derived from symbolic regression accelerating the discovery of new perovskite catalysts.
\newblock {\em Nature communications}, 11(1):3513, 2020.

\bibitem{wilmer2012large}
Christopher~E Wilmer, Michael Leaf, Chang~Yeon Lee, Omar~K Farha, Brad~G Hauser, Joseph~T Hupp, and Randall~Q Snurr.
\newblock Large-scale screening of hypothetical metal--organic frameworks.
\newblock {\em Nature chemistry}, 4(2):83--89, 2012.

\end{thebibliography}

\clearpage
\appendix
\section{Learned rules}\label{sec:appendix:learned_rules}
In this section, we analyse the interpretability of the rules learned by SyMDis for both the Decision Tree and FastLAS variants. Before doing so, Table \ref{tab:desc} presents an overview of the descriptors used from \cite{moosavi2020understanding}. 

\begin{table}[h]
\caption{MOF descriptors used from \cite{moosavi2020understanding}}
\vspace{4pt}
\label{tab:desc}
\centering
\resizebox{0.75\linewidth}{!}{%
\begin{tabular}{@{}ccc@{}}
\toprule
\textbf{Name}           & \textbf{Description}                                               & \textbf{Units}   \\ \midrule
ASA                     & \textit{Accessible surface area to volume ratio}                            & $m^{2}\ cm^{-3}$ \\ \midrule
CellV                   & \textit{Volume of unit cell of MOF}                                         & $\mathring{A}^3$ \\ \midrule
density                 & \textit{Density}                                                            & $g\ cm^{-3}$     \\ \midrule
Df                      & \textit{Diameter of largest free sphere}                                    & $\mathring{A}$   \\ \midrule
Di                      & \textit{Diameter of largest included sphere}                                & $\mathring{A}$   \\ \midrule
Dif                     & \textit{Diameter of largest included sphere along free path }               & $\mathring{A}$   \\ \midrule
NASA                    & \textit{Non-accessible surface area to volume ratio   }                     & $m^2\ cm^{-3}$   \\ \midrule
POAV                    & \textit{Ratio of accessible pore volume to mass}                            & $cm^3\ g^{-1}$   \\ \midrule
POAVF                   & \textit{Fraction of total\_POV\_volumetric that is accessible to probe}     & -                \\ \midrule
PONAV                   & \textit{Ratio of non-accessible pore volue to mass}                         & $cm^3\ g^{-1}$   \\ \midrule
PONAVF                  & \textit{Fraction of total\_POV\_volumetric that is non-accessible to probe} & -                \\ \midrule
total\_SA\_volumetric   & \textit{Surface area to volume ratio regardless of accessibility}           & $m^2\ cm^{-3}$    \\ \midrule
total\_SA\_gravimetric  & \textit{Surface area to mass ratio regardless of accessibility}             & $m^2\ g^{-1}$    \\ \midrule
total\_POV\_volumetric  & \textit{Ratio of pore volume to MOF volume regardless of accessibility}     & -                \\ \midrule
total\_POV\_gravimetric & \textit{Ratio of pore volume to mass regardless of accessibility}           & $cm^3\ g^{-1}$   \\ \bottomrule
\end{tabular}
}
\end{table}

% % performance is DT 33.9904989017/34.1348947775, FastLAS 33.795480417899995/34.1348947775.
% % BW20K Repeat 6
% \begin{figure}[h]
%     \centering
%     \begin{subfigure}[m]{0.46\textwidth}
%         \begin{lstlisting}
% label(excellent) :- CellV > 5827.87, total_SA_gravimetric > 4914.62, Df <= 11.44, density > 0.35.

% label(good) :- CellV > 5827.87, total_SA_gravimetric > 4914.62, Df > 11.44.
% \end{lstlisting}
%     \caption{SyMDis Decision Tree excellent class only}
%     \label{fig:dt_rules}
%     \end{subfigure}
%     \hspace{10pt}
%     \begin{subfigure}[m]{0.41\textwidth}
%         \begin{lstlisting}
% label(excellent) :- CellV >= 26335.50.
% label(good) :- 13.77 <= Dif <= 13.89.
% label(moderate) :- 0.36 <= density <= 0.38.
% label(poor) :- 0.39 <= density <= 0.43.
%     \end{lstlisting}    
%         \caption{SyMDis FastLAS}
%         \label{fig:fastlas_rules}
%     \end{subfigure}
%     \caption{BW20K repeat 6 learned rules}
%     \label{fig:learned_rules}
% \end{figure}

Figure \ref{fig:learned_rules} shows the rules learned for one repeat on the BW20K dataset. During the optimisation for this repeat, the best MOFs discovered were 99.58\% and 99.01\% of the maximum for the Decision Tree and FastLAS variants respectively, which indicates these models both had strong performance. %Both rule sets show that the unit cell volume descriptor is important. The unit cell within the MOF is the structure that is repeated to make the crystal. It's possible that a larger unit cell is likely to have more space, which could influence adsorption \cite{guo2015zeolite}. This provides a reasonable chemical explanation for this descriptor being learned. 
In terms of interpretability, it is immediately clear that the rules learned by the Decision Tree variant are longer than the rules learned by FastLAS, and are therefore more difficult to interpret. This is why we only show the rules learned for the \textit{excellent} and \textit{good} classes for the SyMDis Decision Tree variant in Figure \ref{fig:dt_rules}. To investigate this further, we analyse the interpretability of the rules learned by both SyMDis variants, in terms of the average number of rules learned, and the average length per rule, over the 20 repeats. The intuition is that a shorter set of rules is easier to interpret. Figure \ref{fig:rule_len_analysis} presents the results. As you can see, the rules learned by FastLAS are significantly easier to interpret than those learned by the decision tree. Furthermore, the number of rules, and the average length per rule both increase with larger datasets (CoRE2019 and BW20K) for the Decision Tree variant, whereas the size of the FastLAS rules remains constant regardless of dataset size. This indicates that interpreting the rules learned by the Decision Tree may become more difficult when large datasets are used. In practice, one may wish to prioritise interpretability over performance. For example, in Figure \ref{fig:core2019}, the SyMDis Decision Tree variant outperforms SyMDis FastLAS, but it is more difficult to assess whether the rules learned by the Decision Tree correspond to a reasonable physical and chemical explanation due to their increased length.

In terms of expressivity, in this work the learned rules are propositional and act as a linear classifier. However, one of the benefits of enabling modularity to the symbolic learning component is that higher-order rules could be learned in other tasks using the FastLAS symbolic learner. As FastLAS is based on ASP \cite{gelfond2014knowledge}, it can learn highly expressive first-order rules involving negation, choice, constraints, and rules with multiple answer-sets. This would not be possible with the decision tree SyMDis variant. For the purposes of SyMDis, any set of logical rules are supported, provided they can act as a filter on the database for the next iteration. This is feasible, as each database candidate can simply be evaluated against the learned rules for satisfiability, to decide whether or not the candidate is selected for the next iteration.

\section{Generalisation of learned rules with zero-shot transfer}\label{sec:app:generalisation}
One of the benefits of learning an interpretable set of rules is the ability to transfer to new target datasets, without requiring \textit{any} calls to the expensive objective function. As the datasets used in this paper are from different chemical spaces \cite{moosavi2020understanding}, it is interesting to evaluate whether the rules learned on a source dataset can be transferred to a target. This would indicate the learned rules are general, and possibly reflect common underlying chemical or physical phenomena. It is also clearly of practical interest, as a successful zero-shot transfer would significantly reduce the number of evaluations of the objective function. Note that such a transfer is easier to achieve than with a black-box model, where all descriptors present in the source dataset would have to exist in the target dataset. A set of rules on the other hand, can easily be modified in the case of a partial match, where descriptors could be removed from the rules if they don't exist in the target dataset, or certain descriptors could be transformed if the target dataset had any normalisation or other transformation applied. Also, the rules support manual additions by human experts, if any background knowledge or constraints were required in the target domain. This would be difficult to achieve with a black-box model.

We apply the rules learned at the end of each SyMDis optimisation (i.e., the points with the maximum number of calls to the objective function in Figure \ref{fig:results}) to unseen target datasets. Specifically, we evaluate all combinations of source/target pairs with the ARABG, CoRE2019, and BW20K datasets, except where the ARABG dataset is the target, since this only contains 387 MOFs. We use the learned rules to obtain a predicted class $\hat{c}$ for every material in the target dataset, and take a random sample of 200 materials predicted as \textit{excellent}. We then plot the distribution of working capacities from the sampled MOFs, and compare to 200 random samples. If the rules have generalised, we expect the 200 samples obtained via the \textit{excellent} prediction from the learned rules to have a larger working capacity than the random samples.

\begin{figure}[h]
    \centering
    \includegraphics[width=0.5\textwidth]{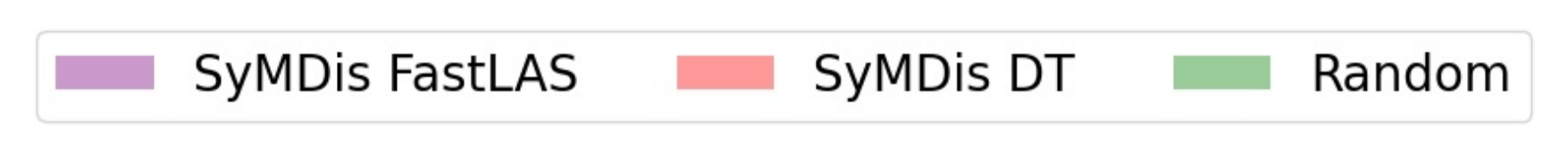}
    \includegraphics[width=1\textwidth]{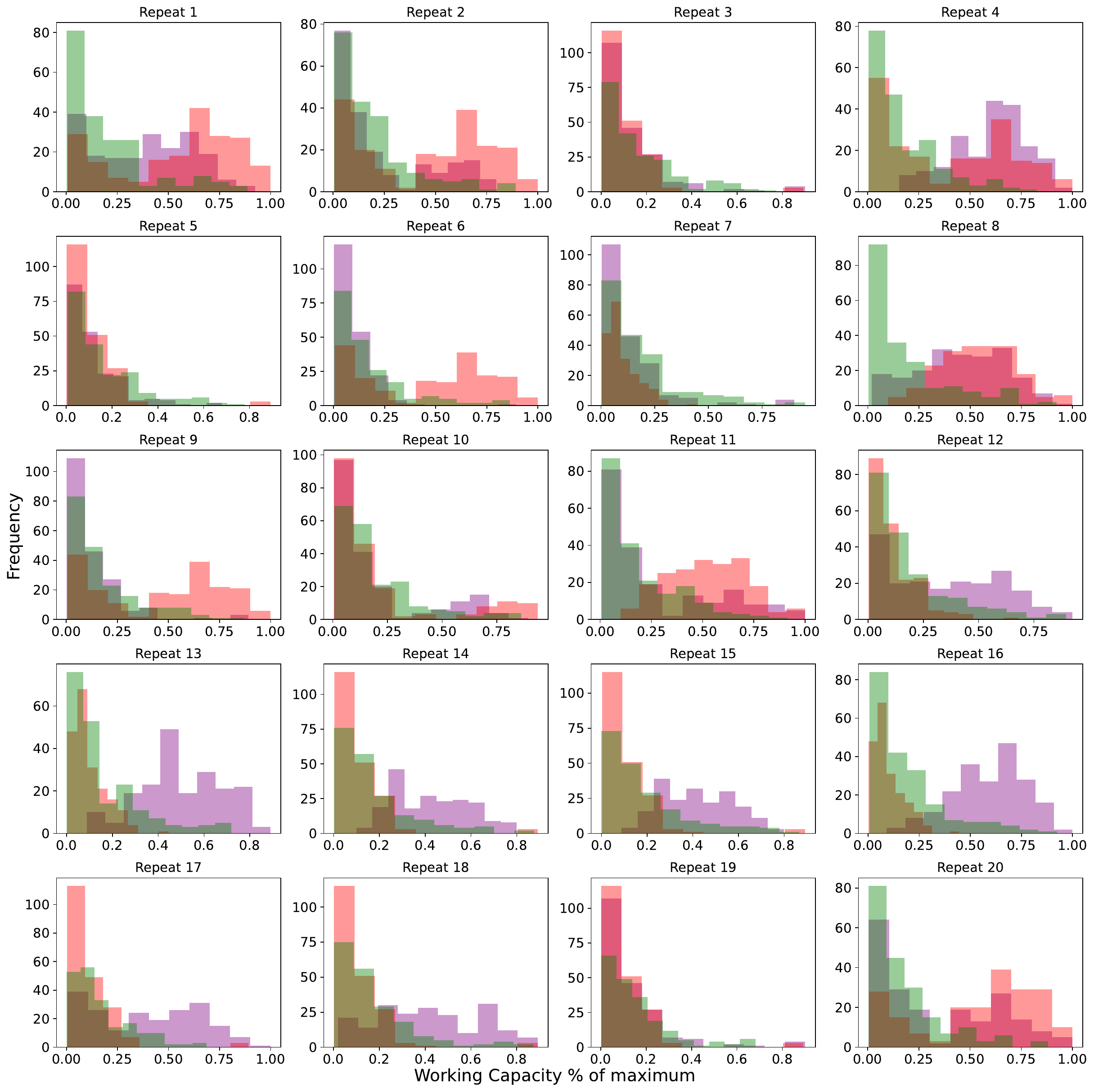}
    \caption{Source = ARABG, target = Core2019}
    \label{fig:arabg_to_core2019}
\end{figure}

\begin{figure}[h]
    \centering
    \includegraphics[width=0.5\textwidth]{figures/gen_legend.pdf}
    \includegraphics[width=1\textwidth]{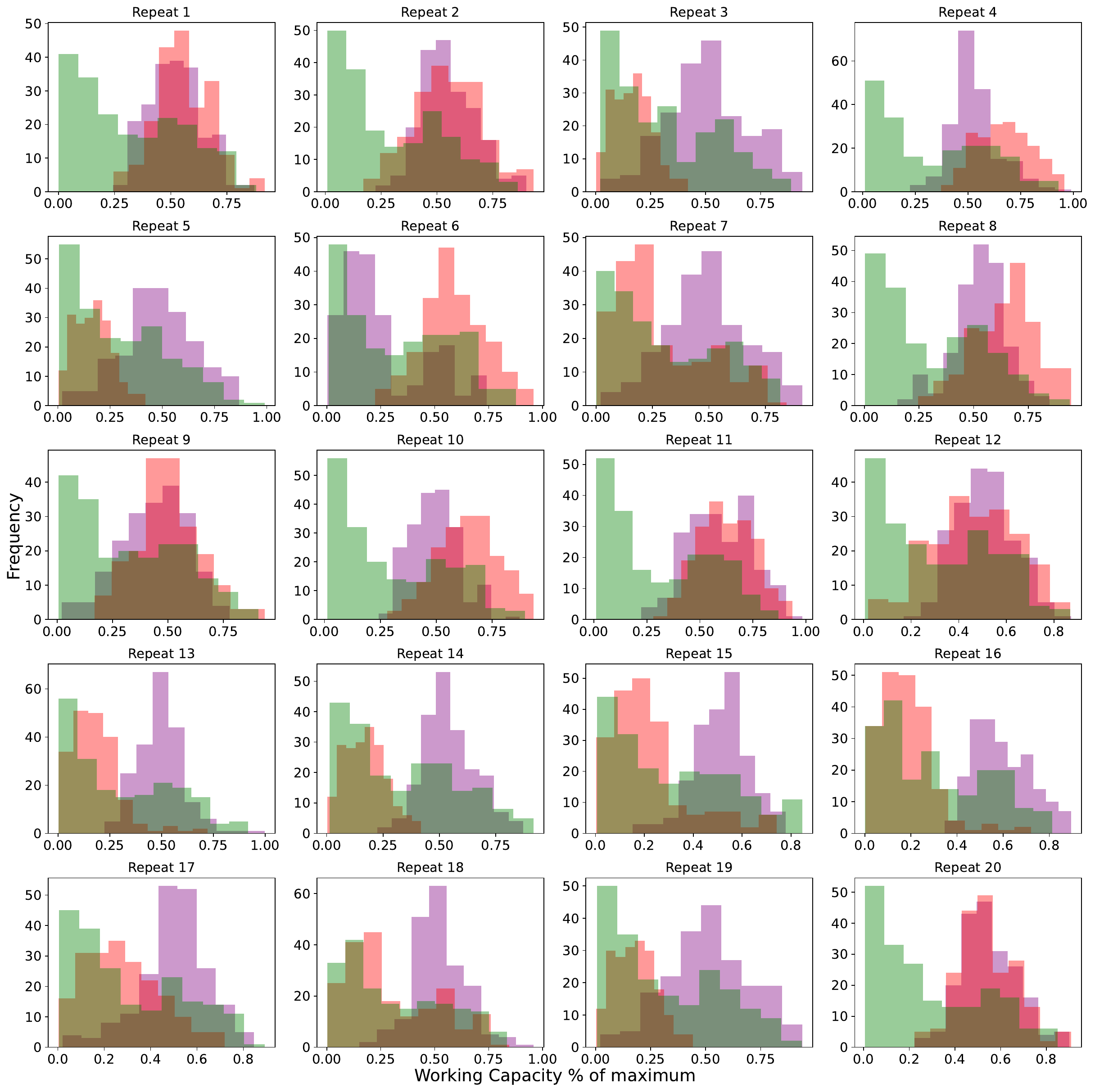}
    \caption{Source = ARABG, target = BW20K}
    \label{fig:arabg_to_bw20k}
\end{figure}

\begin{figure}[h]
    \centering
    \includegraphics[width=0.5\textwidth]{figures/gen_legend.pdf}
    \includegraphics[width=1\textwidth]{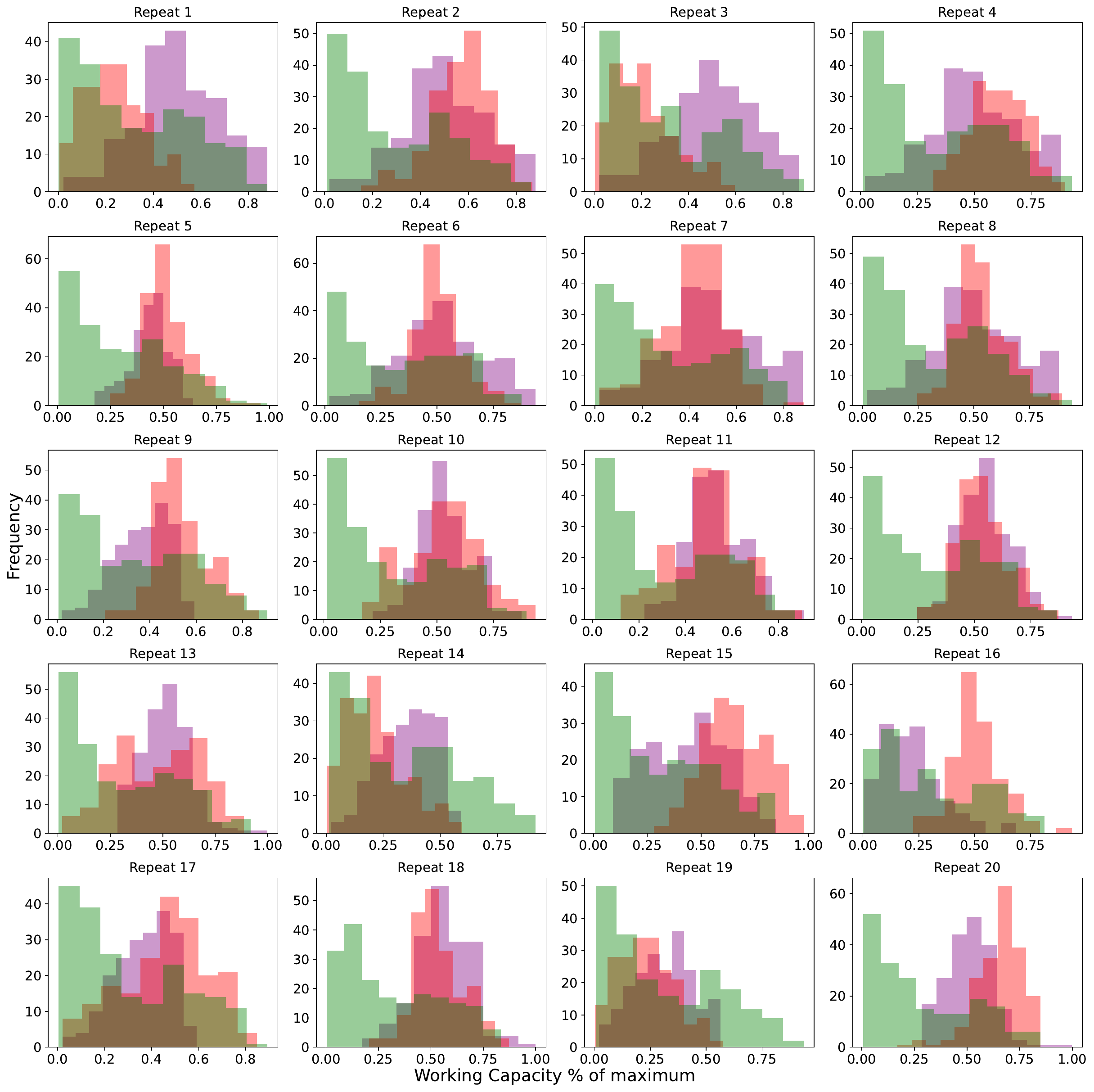}
    \caption{Source = CoRE2019, target = BW20K}
    \label{fig:core2019_to_bw20k}
\end{figure}

\begin{figure}[h]
    \centering
    \includegraphics[width=0.5\textwidth]{figures/gen_legend.pdf}
    \includegraphics[width=1\textwidth]{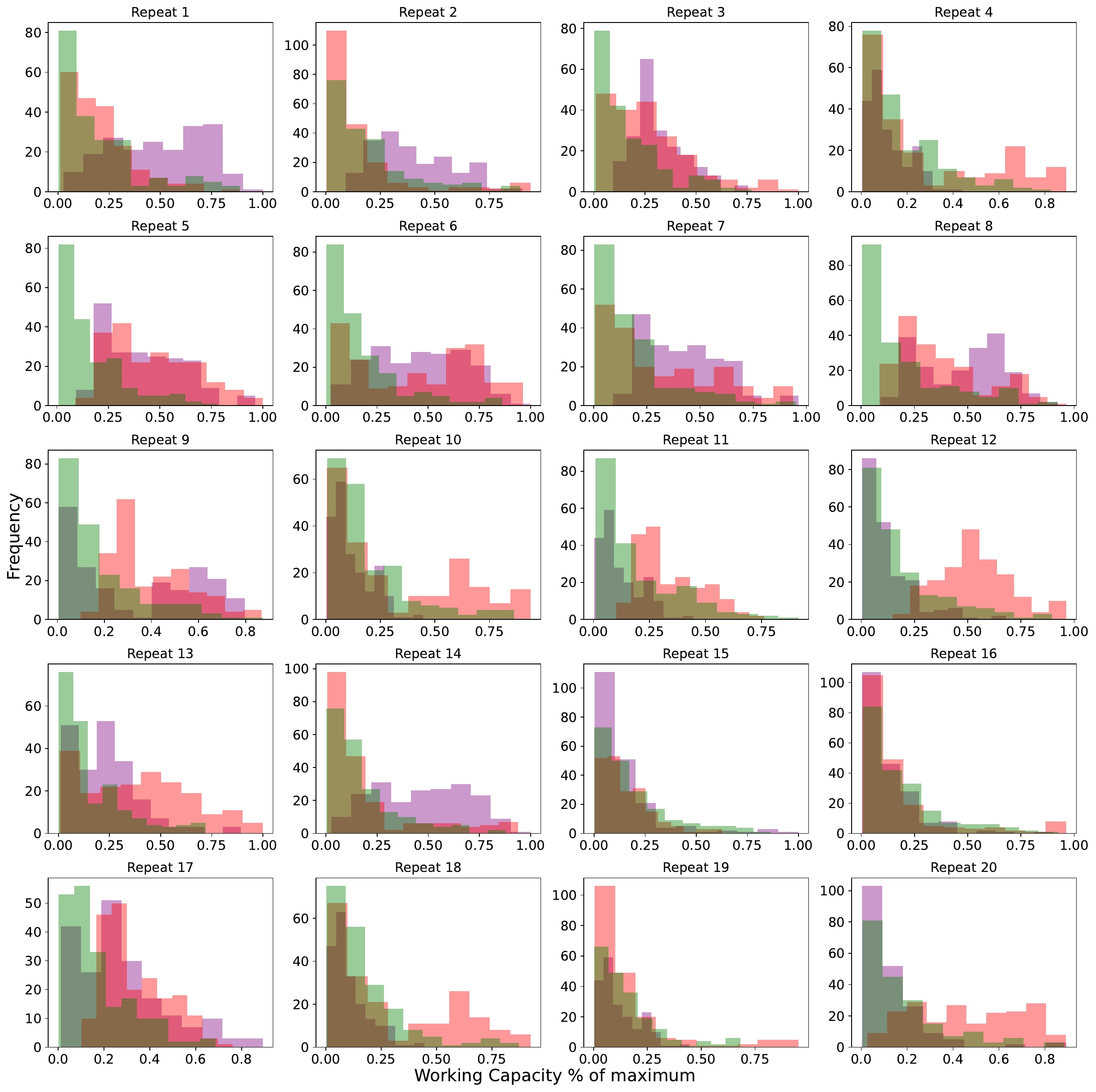}
    \caption{Source = BW20K, target = CoRE2019}
    \label{fig:bw20k_to_core2019}
\end{figure}

Figures \ref{fig:arabg_to_core2019} - \ref{fig:bw20k_to_core2019} show the results for all combinations and all 20 repeats. SyMDis provides a clear benefit in; Figure \ref{fig:arabg_to_core2019} (ARABG to CoRE2019) with 15/20 repeats, Figure \ref{fig:arabg_to_bw20k} (ARABG to BW20K) with all repeats, Figure \ref{fig:core2019_to_bw20k} (CoRE2019 to BW20K) with 18/20 repeats, and in Figure \ref{fig:bw20k_to_core2019} (BW20K to CoRE2019) with 16/20 repeats. This shows the rules learned by SyMDis are able to generalise during a zero-shot transfer and outperform a random selection. In Figures \ref{fig:arabg_to_core2019} (ARABG to CoRE2019) and \ref{fig:arabg_to_bw20k} (ARABG to BW20K), FastLAS provides a benefit compared to the Decision Tree in 7/20, and 10/20 repeats respectively, whereas in other cases the difference is less significant. This is possibly due to FastLAS learning shorter rules, which are more likely to generalise. The rules learned on BW20K do not generalise as well when applied to CoRE2019 (Figure \ref{fig:bw20k_to_core2019}). As BW20K is a \textit{hypothetical} dataset, and CoRE2019 is \textit{experimental} \cite{moosavi2020understanding}, it's likely that some of the rules learned on BW20K may not apply experimentally. Nevertheless, the interpretability of SyMDis enables such analysis and investigation before downstream application, which would be difficult to achieve with a black-box approach. Also, we consider a rule transfer from a hypothetical to an experimental dataset a special case, and the more common use case being a transfer from an experimental to a hypothetical dataset, such as the transfer from CoRE2019 to BW20K in Figure \ref{fig:core2019_to_bw20k}. In this case, the rules have improved generalisation compared to Figure \ref{fig:bw20k_to_core2019}, and help to select high performing \textit{hypothetical} MOFs that can be further validated experimentally.

\end{document}